\documentclass[prc,12pt,superscriptaddress,onecolumn,showpacs,amsmath,amssymb,aps,nofootinbib,floatfix]{revtex4-2}
\usepackage{tikz}
\usetikzlibrary{calc,trees,positioning,arrows,chains,shapes.geometric,  decorations.pathreplacing,decorations.pathmorphing,shapes,  matrix,shapes.symbols}
\usepackage{epsfig}
\usepackage{xcolor}
\usepackage{amsmath}
\usepackage{mathtools,slashed}
\usepackage{amssymb}
\usepackage[colorlinks=true, linkcolor=blue, citecolor=blue, urlcolor=blue]{hyperref}
\usepackage{setspace}
\usepackage[letterpaper,margin=1in]{geometry} 

\newcommand{\be}{\begin{eqnarray}}
\newcommand{\eqn}[1]{Eq.\,(\ref{#1})}

\newcommand{\ee}{\end{eqnarray}}

\def\apj{Astrophys.\ J.}

\def\araa{ARA\&A}
\def\mnras{Mon.\ Not.\ R.\ Astron.\ Soc.}

\def\jhep{J.\ High\ Energ.\ Phys.}

\def\prd{Phys.\ Rev.\ D\;}

\begin{document}
\doublespacing

\title{When Weak Fields Aren’t Weak: \\Post-Newtonian effective theory and the Dark Matter Puzzle}
\author{Marco Galoppo}
\thanks{Corresponding author}
\email{marco.galoppo@pg.canterbury.ac.nz}
\affiliation{School of Physical \& Chemical Sciences, University of Canterbury, Private Bag 4800, Christchurch 8140, New Zealand}
\author{Giorgio Torrieri}
\email{torrieri@unicamp.br}
\affiliation{Universidade Estadual de Campinas - Instituto de Física Gleb Wataghin\\
Rua Sérgio Buarque de Holanda, 777\\
CEP 13083-859 - Campinas - São Paulo - Brasil
}
\noindent \textnormal{Essay written for the Gravity Research Foundation 2026 Awards for Essays on Gravitation.\\ Submitted on the 8th march 2026}

\begin{abstract}
Post-Newtonian theory is considered a reliable effective expansion of General Relativity in the weak-field and slow-motion limit. We argue that such a belief is misplaced. In generic many-body relativistic dynamics, the absence of globally conserved charges in the region of interest and non-integrability can drive strong sensitivity to angular-momentum exchange across inhomogeneous curvature, invalidating naive power counting in an effective theory expansion. Building on general lessons from effective field theory, we derive an explicit breakdown criterion that delineates when post-Newtonian truncations become unreliable despite small local potentials and velocities. This supplies a controlled systematic for weak-field mass inference, relevant to the dark matter puzzle in astrophysics and cosmology. 
\end{abstract} 
\maketitle
\newpage

General Relativity (GR) is the currently accepted theory of gravity, describing the gravitational interaction in terms of spacetime geometry~\cite{gravitation}. The fundamental degrees of freedom of the theory are encoded within the spacetime metric, whilst the relatively simple local gravitational action---namely, the Einstein--Hilbert action---is restricted by diffeomorphism invariance, and expressed in terms of the Ricci scalar together with a cosmological constant (and matter fields). The resulting field equations are nonlinear and constrained, so analytic solutions are scarce outside highly symmetric settings. Additionally, the same gauge symmetry that underlies the theory’s elegance also complicates both approximation schemes and numerical evolution, which must respect constraint propagation and coordinate freedom.

For these reasons, most practical astrophysical calculations de facto adopt Newtonian gravity, in which the interaction is mediated by a scalar potential sourced by mass density, and general relativistic effects can be incorporated as perturbative corrections. In particular, the leading empirical challenge in large-scale gravitational physics---the Dark Matter (DM) problem~\cite{Bertone_2005}---is often examined in essentially Newtonian terms: galaxy rotation curves, the dynamics of galaxy groups and clusters, and related observables are typically modelled using Newtonian (or post-Newtonian) dynamics. Even many alternatives to particle DM, such as modified gravity theories, are frequently introduced at the level of non-relativistic phenomenology, with a fully relativistic completion treated as a subsequent step~\cite{mond}. 

An outsider might object that one should not expect a complete explanation of astrophysical observations from a theory superseded more than a century ago: perhaps the DM puzzle is simply an artefact of insisting on Newtonian gravity where GR should be used. Part of what makes GR compelling, however, is precisely that it reproduces Newtonian gravity in the appropriate regime. Writing $g_{\mu\nu}\simeq \eta_{\mu\nu}+h_{\mu\nu}$, with $\eta_{\mu\nu}$ the Minkowski metric and $h_{\mu\nu}$ a small perturbation, and assuming slow motion $v\ll c$, the field equations and equations of motion admit a controlled asymptotic expansion organized in powers of $v/c$ and of $h_{\mu\nu}$. At leading order one recovers Newtonian gravity, with successive post-Newtonian corrections scaling as $\mathcal{O}((v/c)^n)$ and $\mathcal{O}(h^n)$~\cite{Poisson_2014,Clifford_2014}. In the astrophysical settings where Newtonian modelling is commonly employed, these higher-order terms are typically assumed to be negligible.

This viewpoint is an instance of the broader effective-theory philosophy: when a problem exhibits a separation of scales (here set by small deviations from local flatness and small velocities), the dynamics can be organized as an expansion in the corresponding small parameters, and only the first few terms should materially affect observables. Effective descriptions of this kind have been remarkably successful~\cite{eftintro}, and the post-Newtonian expansion has been formulated explicitly in this language as well~\cite{Porto_2016}.

However, a puzzling issue remains: there exist toy models that are simple enough to admit fully general-relativistic solutions, yet realistic enough to mimic basic features of rotating galaxies~\cite{Balasin_2008,Astesiano_2022a,Astesiano_2022L,Beordo_2024,Galoppo_2025,Astesiano_2025}. Strikingly, these models fail to reproduce the Newtonian limit even in regimes where such a limit would be the expected outcome. As argued in \cite{Galoppo_2025,Astesiano_2025}, taking the weak-field expansion of the corresponding GR solutions appears to generate additional leading-order contributions that are absent from the standard post-Newtonian expansion about a Minkowski background. This would then be in direct contrast to the effective field theory paradigm. Yet because the discrepancies are highly solution-dependent and the models remain idealized, it has been difficult to isolate (i) precisely how the effective description breaks down, (ii) what the corresponding ``missing'' operators in an effective Lagrangian should look like in general, and (iii) under what physical conditions they become non-negligible. This combination of ambiguity and idealization has led many to dismiss these examples as pathological or nongeneric, leaving the post-Newtonian effective theory largely unchallenged and regarded as fully reliable~\cite{Ciotti_2022}.

In this essay, we propose a different explanation: precisely because of the mathematical structure and symmetries of GR, a naive Taylor-like expansion in small corrections can break down even when the nominal expansion parameters are tiny. We examine the situations in which such a breakdown is expected, and relate them to the regimes where dark matter is commonly invoked to explain astrophysical observations.

The potential flaw in expanding in terms of perturbative parameters is not merely quantitative, but structural: it concerns how conservation laws and integrability properties match between the Newtonian and general relativistic descriptions. A key consequence of general covariance is that conservation statements of the form $\partial_\mu J^{\mu\cdots}=0$, where $J^{\mu\cdots}$ is a current constructed from the fields, are generally meaningful only locally. On the scales where tidal effects are relevant---that is, where spacetime curvature cannot be neglected---there need not exist a corresponding globally conserved quantity obtained by integrating such a current over an extended region.

This observation singles out angular momentum conservation, both because angular momentum is inherently non-local and because its conservation is a (hidden) key organizing principle in Newtonian dynamics and in standard DM reasoning, where the original mass-discrepancy arguments rely heavily on virialized, quasi-cyclic motion and the virial theorem. This is already conceptually troubling from an effective field theory viewpoint~\cite{Donoghue_2009}, where one typically assumes that the symmetries realised by the effective theory are inherited from the underlying theory, rather than introducing genuinely new exact symmetries absent in the fundamental description. 

General Relativity possesses a deep local gauge symmetry---diffeomorphism invariance---but, in generic curved spacetimes, global spacetime symmetries of the sort that underwrite conserved charges need not exist~\cite{noglobal1}. This absence of exact global symmetries is a structural feature of GR, and is often argued to be even more stringent in the quantum-gravity context~\cite{noglobal2,noglobal3}. By contrast, Newtonian gravity is formulated on a fixed background with global invariance properties, yielding a series of exact conserved charges; e.g., angular momentum conservation follows from the global rotational symmetry of the Newtonian action.

Here, the point is that conservation laws are, in a sense, mathematically and physically special: they are what render a system integrable~\cite{Goldstein_2002}, allowing a reduction to action--angle variables and the quasi-periodic motions that underlie results such as the virial theorem. Hence, any departure from such conservation laws will inevitably introduce non-integrable perturbations within the system.

The Kolmogorov--Arnold--Moser (KAM) theorem~\cite{Goldstein_2002} shows that integrability is stable under sufficiently small non-integrable perturbations, but the required notion of ``sufficiently small'' is inversely proportional to the number of degrees of freedom within the system. Consequently, a system with many interacting components and substantial angular-momentum exchange can lose effective integrability due to general relativistic corrections even in the presence of weak fields and for small velocities. In that regime, small relativistic terms can qualitatively alter the long-time dynamics: nearby trajectories separate rapidly, and the evolution develops sensitivity that is not captured by a uniform expansion in the nominal small parameters, despite their instantaneous smallness.

Motivated by these considerations, we now introduce a quantitative breakdown diagnostic~\cite{us} designed to be sensitive to precisely the two caveats above: the (lack of) global conserved charges and the loss of effective integrability in large many-body systems. The quantity should (i) vanish in the flat/vacuum limit, (ii) be sensitive to angular-momentum transport through an extended region, (iii) be sensitive to spacetime curvature, and (iv) be a dimensionless scalar, since it is intended to function as an expansion/validity parameter.

For angular momentum, it is convenient to work with the angular-momentum current (density) $J^{\mu\nu\alpha}(x):=x^\mu T^{\nu\alpha}(x)-x^\nu T^{\mu\alpha}(x)$, built from the stress-energy tensor $T^{\mu\nu}$ and locally Minkowski coordinates. The associated charge on a spacelike hypersurface $\Sigma=(t(\vec{x}),\vec{x})$ is then $J^{\mu\nu}[\Sigma]:=\int_{\Sigma} J^{\mu\nu\alpha}\, \mathrm{d}^3\Sigma_\alpha$, where $\mathrm{d}^3\Sigma_\alpha=\epsilon_{\alpha \beta \mu \nu} \frac{\partial \Sigma^\beta}{\partial x_1} \frac{\partial \Sigma^\mu}{\partial x_2}\frac{\partial \Sigma^\nu}{\partial x_3} dx_1 dx_2 dx_3$ is the Lorentz-covariant directed hypersurface 3-volume element.

The spacetime curvature enters instead through the Riemann tensor $R_{\mu\nu\alpha\beta}$. Since violations of would-be global conservation laws are controlled by tidal effects, any such diagnostic must couple angular-momentum flow to curvature. The simplest scalar with the desired ``doubly extensive'' (pairwise, nonlocal) structure is a double volume integral $\tilde{\alpha}$ defined by
\begin{equation}
\tilde{\alpha}=G\int_{\Sigma_x} \mathrm{d}^3\Sigma_{\xi'}(x)\int_{\Sigma_y} \mathrm{d}^3\Sigma_{\zeta}(y)\,\mathcal{R}_{\alpha'\beta'\gamma\delta}(x,y)\,J^{\alpha'\beta'\xi'}(x)\,J^{\gamma\delta\zeta}(y)\;.
\end{equation}
Here, $\mathcal{R}_{\alpha'\beta'\gamma\delta}(x,y)$ denotes the appropriate curvature bi-tensor (i.e., the Riemann tensor evaluated at $x$ with its indices parallel-transported to $y$ along the connecting geodesic), so that the contraction in $\tilde{\alpha}$ is covariant. The explicit factor of $G$ both fixes the dimensionality of the diagnostic parameter and guarantees the correct decoupling limit with respect to gravity, namely $\tilde{\alpha}\to 0$ as $G\to 0$.

Schematically, one therefore expects a rough scaling as 
\begin{equation}
    \label{scaling}
\tilde{\alpha}\sim G\,\langle R\rangle\,\langle J\rangle^{2}\,\langle V\rangle^{2}
\end{equation}
where $R$ is the Ricci scalar, $J$ the angular momentum and $V$ the volume, reflecting the double integration over pairs of volume elements and the dependence on both angular momentum and curvature. We call $\tilde{\alpha}$ ``doubly extensive'' because it sums interaction-like contributions between each volume element and every other one. Interestingly, analogous double-extensive quantities appear, for example, in mean-field descriptions of van der Waals gases and in Quantum ChromoDynamics (QCD) as we discuss further on.

Finally, for astrophysical coarse graining, it is natural to normalize dimensionless estimates using the scale $G M_\odot^2$, treating a star (with characteristic unit mass $M_\odot$) as an elementary constituent degree of freedom of many astrophysical systems. Operationally, one fixes a coarse-grained description of the system (analytic density/velocity profiles for idealized systems, or voxelized density--velocity reconstructions for data-driven ones), constructs an effective energy-momentum tensor, and evaluates the curvature entering $\tilde{\alpha}$ at leading post-Newtonian order via the Newtonian potential's tidal tensor. Here, we note that using the leading post-Newtonian estimate for the curvature remains justified, as $\tilde{\alpha}$ flags only when post-Newtonian dynamics becomes unreliable due to cumulative nonlocal angular-momentum exchange across inhomogeneous curvature rather than from large pointwise curvature corrections. One then can compute $\tilde{\alpha}$ as a double volume integral (or its discretized grid analogue) coupling this curvature to the relativistic angular-momentum density. For survey-level estimates this reduces to a scaling in terms of independently inferred $\langle R\rangle$, $\langle J\rangle$, and $\langle V\rangle$ as of \eqn{scaling}. The results of some of these estimates are summarized in Table~\ref{tab:1} (see \cite{us} for further technical details).

What is tantalizing is that the diagnostic parameter introduced here is small precisely in regimes where the post-Newtonian expansion is quantitatively well tested: $\tilde{\alpha}$ is negligible for both stellar systems and pulsar binaries. Conversely, $\tilde{\alpha}$ becomes large in many of the settings where a mass discrepancy is usually inferred and DM is invoked, including galactic rotation curves and the dynamics of galaxy clusters (and, plausibly, systems such as the Bullet Cluster). 
\\
\begin{table*}[htb!] \centering \setlength{\tabcolsep}{10pt} 
\renewcommand{\arraystretch}{1.1} 
\begin{tabular}{lllll} 
\hline\hline
Stellar binary & Pulsar binary & Elliptical galaxy & Disc galaxy & Laniakea \\
\hline
$10^{-9}$ & $10^{-5}$ & $10^{9}$ & $10^{10}$ & $10^{26}$ \\
\hline\hline
\end{tabular} 
\caption{Estimates of $\tilde{\alpha}$ for a series of astrophysical systems}\label{tab:1} 
\end{table*}

One important exception is the CMB anisotropy spectrum, which is well fit by a $\Lambda$CDM inflationary framework in which angular momentum does not enter in any direct way~\cite{Planck_2016}. This caveat would be weakened, however, if cosmological-scale magnetic fields are confirmed to be present at dynamically relevant levels during inflation or its aftermath, as suggested in some scenarios~\cite{Matarrese_2005,Durrer_2013}; in that case, vorticity and angular-momentum transport could again become part of the effective description.

The findings presented here are certainly not conclusive, but they motivate several concrete next steps. On the theory side, a dedicated analysis using an $N$-body or numerical hydrodynamics code that incorporates full GR~\cite{rezzolla} could ascertain the admixture of dark matter required to model galactic dynamics in light of the breakdown of post-Newtonian expansion. In particular such simulations could test whether deviations from many-body Newtonian gravity correlate with $\tilde{\alpha}$ in a genuinely non-perturbative manner. In such a framework, the ingredients entering $\tilde{\alpha}$ can, in principle, be varied systematically (e.g.\ total angular momentum, degree-of-freedom count/coarse graining, and curvature inhomogeneity), allowing one to map out the boundary of post-Newtonian reliability. Observationally, large-scale lensing and dynamics surveys (DESI, Euclid, LSST~\cite{synergy}) can be used to test whether $\tilde{\alpha}$ functions as a scaling variable tracking inferred DM content as of \eqn{scaling}. In particular, ultra-faint dwarf galaxies~\cite{Simon_2019} provide an attractive probe: within a relatively narrow baryonic mass range, they exhibit large morphological diversity, offering leverage to disentangle mass from angular-momentum/structure systematics.

The discussion above has a suggestive, if not exact, analogue in Yang--Mills theory, another field theory in which deep local symmetries enforce nonlinearity and effective nonlocality. In non-Abelian gauge theories, even the notion of ``weak fields'' is subtle because the gauge can only be fixed locally, and a given covariant gauge condition admits infinitely many gauge-equivalent configurations (``Gribov copies'') related nonlinearly to the gauge potential. This is the Gribov problem~\cite{Gribov_1978}. Only when the characteristic wavenumbers are high enough that gauge artefacts average out---so that the relevant fields are smooth on the scales of interest and can be effectively linearized---does a naive weak-field expansion become reliable (this intuition underlies asymptotic freedom). Beyond that regime, a controlled expansion must instead be formulated in terms of quantities that are insensitive to Gribov redundancies. The canonical examples are Wilson loops~\cite{Wilson_1974}, which are intrinsically nonlocal and nonlinearly related to the fundamental gauge potential. In particular, they also encode the field contribution to the angular momentum carried by coloured charges. This illustrates a general moral: local gauge symmetries---tied to conserved charges through Gauss constraints~\cite{gauss}---can make seemingly straightforward expansions and truncations unreliable when global structure and many-body dynamics are essential.

From this perspective, $\tilde{\alpha}$ may be viewed as loosely analogous to an interaction measure between Wilson loops. Indeed, correlations of Wilson loops provide an order parameter-like diagnostic distinguishing weak- and strong-coupling behaviour in gauge theory~\cite{Wilson_1974}. While GR is profoundly different from Yang--Mills theory---most notably, confinement (often argued to be connected to Gribov ambiguities~\cite{Dudal_2008}) has no direct gravitational analogue---the mathematical parallel is suggestive: local gauge symmetry can force the physically meaningful ``expansion variables'' to be nonlocal and nonlinear. This motivates examining the regime of interest here (many degrees of freedom with large relative angular momentum) using Wilson-loop-inspired techniques, where the effective Lagrangian would become acquire terms in powers of $\tilde{\alpha}$. 
Such methods have been applied to gravity~\cite{White_2011,Bonocore_2022}, albeit in a very different kinematic and conceptual regime than the one considered here.  $\tilde{\alpha}$ is related to the expectation value of a correlator between two spacelike Wilson loops ''large'' in configuration space ($G R^{-2} \rightarrow 0$) coupled to ''many'' sources (the degrees of freedom) and ''high'' angular momentum ($N,L\rightarrow \infty$). Our conjecture is basically that the behavior of this expectation value is not suppressed by a large power but is in fact close to leading ($\sim$ Post-Newtonian) order, and the methods in \cite{White_2011,Bonocore_2022} might be used to test it.

In conclusion, we have examined the applicability of the post-Newtonian expansion and argued that it can fail even in regimes where the nominal relativistic corrections appear parametrically small. A particularly susceptible setting is a many-body system with substantial relative angular-momentum exchange among its constituents across inhomogeneous curvature. The underlying obstruction is structural rather than merely quantitative: in generic spacetimes, GR does not furnish the global symmetries required for exact conserved angular-momentum charges, and the associated loss of integrability means that even weak non-integrable perturbations can qualitatively alter long-time evolution, especially as the number of degrees of freedom grows. We proposed a quantitative diagnostic capturing these effects and found that it naturally distinguishes regimes where post-Newtonian reasoning is well tested from regimes in which mass discrepancies are typically inferred. If borne out by dedicated relativistic simulations and observational scaling tests, this mechanism would provide a principled systematic for weak-field mass inference, with direct implications for the dark-matter problem 
and, more generally, would teach us something fundamentally new about how gravity operates in genuinely many-body systems.

\bibliography{references.bib}
\end{document}